\documentstyle[epsfig]{aipproc}

\begin{document}
\title{A New Approach to Mass Measurements of UHECR:
       Horizontal Air Showers}

\author{E. Zas}
\address{Departamento de F\'\i sica de Part\'\i culas,\\
Universidade de Santiago de Compostela, E-15706 Santiago, Spain.\\
zas@fpaxp1.usc.es}

\maketitle

\begin{abstract}

The role of Horizontal Showers induced by cosmic rays is discussed in 
detail. A new approach to the calculation of the muon component in 
horizontal air showers induced by protons, heavier nuclei or photons 
is presented which allows a simple analytical evaluation of the muon 
density profiles at ground level. The results of the first application 
of these results to horizontal air showers detected at the Haverah Park 
Array by the recently started {\sl Leeds-Santiago collaboration}, leading 
to important restrictions on composition at ultra high energies, 
are reported. 

\end{abstract}

\section*{Introduction}
%
%
%
%

Inclined showers were observed in the 1960's by a number of experiments  
\cite{Tokyo,Kiel,Durham,Andrews} and it was immediately realized that 
they were different from the ordinary vertical showers and they had to 
be due to penetrating particles. At high zenith angles the 
slant atmospheric depth to ground level is enough to absorb the early part 
of the shower that follows from the standard cascading interactions, 
both of electromagnetic and hadronic type. 
Only penetrating particles such as muons and neutrinos can traverse the 
atmosphere at high zeniths to reach the ground or to induce secondary 
showers deep in the atmosphere and close to an air shower detector. 

The idea that it may be possible to detect high energy neutrinos produced 
both in the atmosphere or away from the earth was put forward by Markov in 
1960 \cite{markov}. Horizontal Air Showers (HAS) were at the end of the decade 
suggested as a possible means \cite{berezinskii} but it was soon realized 
that the inclined showers that had been detected were more likely due 
to secondary interactions of the high energy muons that were produced 
in ordinary cosmic ray showers at the top layers of the atmosphere. 
Indeed inclined shower rates in the shower size range between $10^2$ and 
$10^5$ particles have been shown to be consistent with atmospheric muon 
bremsstrahlung \cite{Kiraly,tokiobremss,hsvum}, which is the 
dominant mechanism for HAS. 

From the early days of air shower measurements until now inclined showers 
have been present in many experiments \cite{Antonov} and have given a 
great deal of information. The Haverah Park array 
detected very inclined showers of the highest energies which were 
interpreted as muons from "ordinary" cosmic rays 
(presumably protons or nuclei) \cite{Hillas}. 
In the 1980's it became apparent that one of the interests of 
HAS was the identification of the prompt muon component in the atmospheric 
muon spectrum, an issue that is not yet settled and bounds on muon-poor 
HAS were established \cite{nagano86}. The non-observation of horizontal or 
upcoming air showers by the fluorescence technique gave a limit to the 
diffuse high energy neutrino flux \cite{baltrusaitis}. 

Later on in the 1990's the prospects for high neutrino detection became 
a reality with a number of projects under planning or construction 
and intense theoretical activity resulted in various neutrino production 
mechanisms being proposed and flux predictions estimated \cite{nureview}. 
Bounds on HAS played a significant role constraining 
diffuse neutrino flux predictions from three different mechanisms:  
For the pioneering calculation of the diffuse neutrino flux expected 
from proton acceleration in Active Galactic Nuclei (AGN) by Stecker 
et al. \cite{stecker91}. 
For "top-down" scenarios which have been discussed at 
length by other speakers in this conference, namely the annihilation 
of a class of topological defects, superconducting cosmic strings \cite{Bhattacharjee}. Lastly for "neutrino messenger" models in which 
the Ultra High Energy Cosmic Ray (UHECR) are locally produced by an 
extragalactic flux of high energy neutrinos that interacts with  
massive relic neutrinos clustering around our galactic halo, which 
have also been discussed at this conference \cite{Weiler,Fargion}. 
The muon-poor HAS bound from AKENO \cite{nagano86} has been used to 
constrain both proton acceleration 
in AGN \cite{hznubound} and the annihilation of superconducting cosmic 
strings \cite{BlancoPRL} and the Fly's Eye bound \cite{baltrusaitis} has 
been used to constrain neutrino messenger models \cite{BlancoPRD}. 

Although air shower arrays played an important role in 
neutrino astronomy, the limits on diffuse neutrino fluxes suggested 
that at least in the 10 TeV to 10 PeV band these detectors were more likely 
to play a role in establishing the prompt muon component than in actually 
detecting high energy neutrinos\cite{hsvum}. The atmospheric muon spectrum 
however decreases rapidly as the muon energy rises and as a result HAS 
induced by muon bremsstrahlung become a small background for neutrino 
detection in the upper energy region of the spectrum. 
Detectors aiming at the EeV range of the cosmic ray spectrum and above 
can thus place important constraints on the UHE end of the 
not yet established neutrino spectrum. (Incidentally this is the case for 
the two ambitious  projects to detect Fluorescence and \v Cerenkov light 
from satellites presented in this conference \cite{klypve,euso}.) 
It was shown that the Pierre Auger detector \cite{Auger}, also 
discussed by other 
speakers at this conference, has a large effective volume to neutrino 
detection for zenith angles above $60^{\circ}$ \cite{venice,Capelle}. 
This capability, which nicely connects the UHECR observation to the UHE 
neutrino searches, was clearly dependent on the ability to separate 
those hypothetical neutrino induced showers at high zeniths from a 
background of inclined showers induced by protons, nuclei or photons. 

The many difficulties involved in both detecting and analysing 
inclined showers with particle arrays 
have prevented their systematic use in the past. 
At detection level shower arrays using scintillator detectors were 
usually placed in the horizontal plane making them less efficient 
for horizontal incidence. Moreover the typical array, also lying in a 
horizontal plane, presents a much reduced sampling area for 
near horizontal incidence. 
At the analysis level, when these showers are produced by nucleons or photons, 
the geomagnetic distortions of the particle density profiles 
prevent the use of conventional approach to the study of 
Extensive Air Showers (EAS) such as using the standard measurements of 
particle density at a fixed distance to shower axis (such as $\rho (600)$). 
Data analysis requires comparisons of measurements and expected particle 
density profiles. 
The brute force simulation of these shower profiles would demand 
very lengthy runs and much storage space which would make the data analysis 
extremely cumbersome. This approach was possibly too complicated to be 
justified when the first HAS were being recorded, particularly as shower 
models were quite primitive. 

Understanding the cosmic ray background to neutrino detection by large 
air shower arrays was the original motivation of the study that is 
described in this article. 
The approach of using existing data from the Haverah Park array was 
proposed as a possible way to perform this study which had the clear 
advantage that new ideas about HAS could be immediately tested with 
real data.  Haverah Park data had other specific advantages. 
The water \v Cerenkov detectors 
involved were rather deep and large which makes them more 
suitable for detecting horizontal events than scintillators. 
The signal deposited in them by charged particles is proportional 
to their tracklength so that muons that typically traverse the whole 
tank produce larger signals than photons or electrons. 
Muons constitute the dominant part at ground level of a photon, 
proton or nucleus induced shower at high zenith angles so again 
water \v Cerenkov tanks are at an advantage when detecting inclined 
showers. 
Lastly the Haverah Park array can be considered as a kind of prototype 
of the Auger observatory now under construction in Argentina. 
The proposal led to the Leeds-Santiago collaboration which has recently 
published the first analysis of inclined showers for zeniths above 
$60^{\circ}$ induced by cosmic rays at energies above $10^{19}$~eV 
\cite{avePRL}. 

The ability to analyse inclined showers above $60^{\circ}$ induced by 
nucleons or photons essentially doubles the acceptance of any Air Shower 
array and opens a part of the 
sky that was previously inaccessible to the detector, besides establishing 
the background for neutrino detection. It was a pleasant surprise 
to find that on top of these obvious advantages these showers 
really provide a new tool for UHECR interpretation because 
they are probing muons of significantly higher energies than vertical 
showers. This tool has been shown to be very sensitive to composition 
when combined with vertical shower measurements and relevant conclusions 
have already been obtained from the analysis of the HAS observed at 
Haverah Park. 

\section*{Azimuthal Symmetry in Air Showers}

The standard way to analyse EAS assumes circular symmetry for the 
average particle densities at ground level. There is little doubt that 
this is an accurate symmetry for vertical showers in the atmosphere in 
the absence of geomagnetic effects. Indeed the symmetry principle lies 
behind the standard method for estimating the primary energy of an air 
shower through the relation between primary energy and the particle 
density at a given distance to shower axis, usually 600~m ($\rho (600)$) 
which was demonstrated to be fairly independent of shower fluctuations 
and composition \cite{hillasrho}.  
Even the existence of such a parameter obviously relies on an implicit 
assumption about circular symmetry. 

However it was recently pointed out that deviations from symmetry exist 
and that they become important for correctly estimating the primary energy 
of events particularly for zeniths above $50^{\circ}$ 
\cite{Antonov98,Ivanov}. 
The circular symmetry is broken by the magnetic field and 
for inclined showers also by the density gradient in the atmosphere. 
Clearly particles having transverse (to the shower axis) momenta 
$p_{\perp}$ pointing upwards will develop showers in a thinner 
density atmosphere than those with $p_{\perp}$ pointing downwards. 
Neither of these effects is however very important at low zenith angles.   
The gradient induced asymmetry is expected to be small at all angles 
in the plane perpendicular to shower axis. 
The lateral spread of a shower, of order the Moli\`ere radius 
(10~g~cm$^{-2}$), corresponds to very small differences in matter 
depth travelled by particles in the upper and lower sides of the 
shower plane when compared to the total travelled depth. 

Inclined showers are different from vertical showers in that the 
hadronic and electromagnetic parts of the shower have been completely 
developed and absorbed before reaching the ground. There is 
however a penetrating component of the shower mostly muons and 
neutrinos from pion decays that reaches the ground. 
As a result the muons in the front are 
produced rather high in the atmosphere, at sites which are physically 
rather distant from the observation plane at ground level. 
The approach chosen for the study of inclined showers consists of 
eliminating the effect of the Earth's magnetic field. The study of 
showers in the absence of the earth's magnetic field 
maintains the circular symmetry and hence has enormous advantages 
for the study of the relations between the particle densities at ground 
and all cosmic ray relevant variables such as primary energy, 
zenith angle, composition and extrapolations of the relevant interactions. 
If the effects of the magnetic field can be implemented {\sl  a posteriori} 
that is once the lateral structure of the shower is known, the process 
of calculating particle densities at ground level would become a great 
deal simpler. This has enormous advantages at the levels of event 
simulation, of understanding the physical origin of the inclined shower 
features and of fitting experimental results to theory. 

\section*{Muon Showers}

\begin{figure}[b!] 
\centerline{\epsfig{file=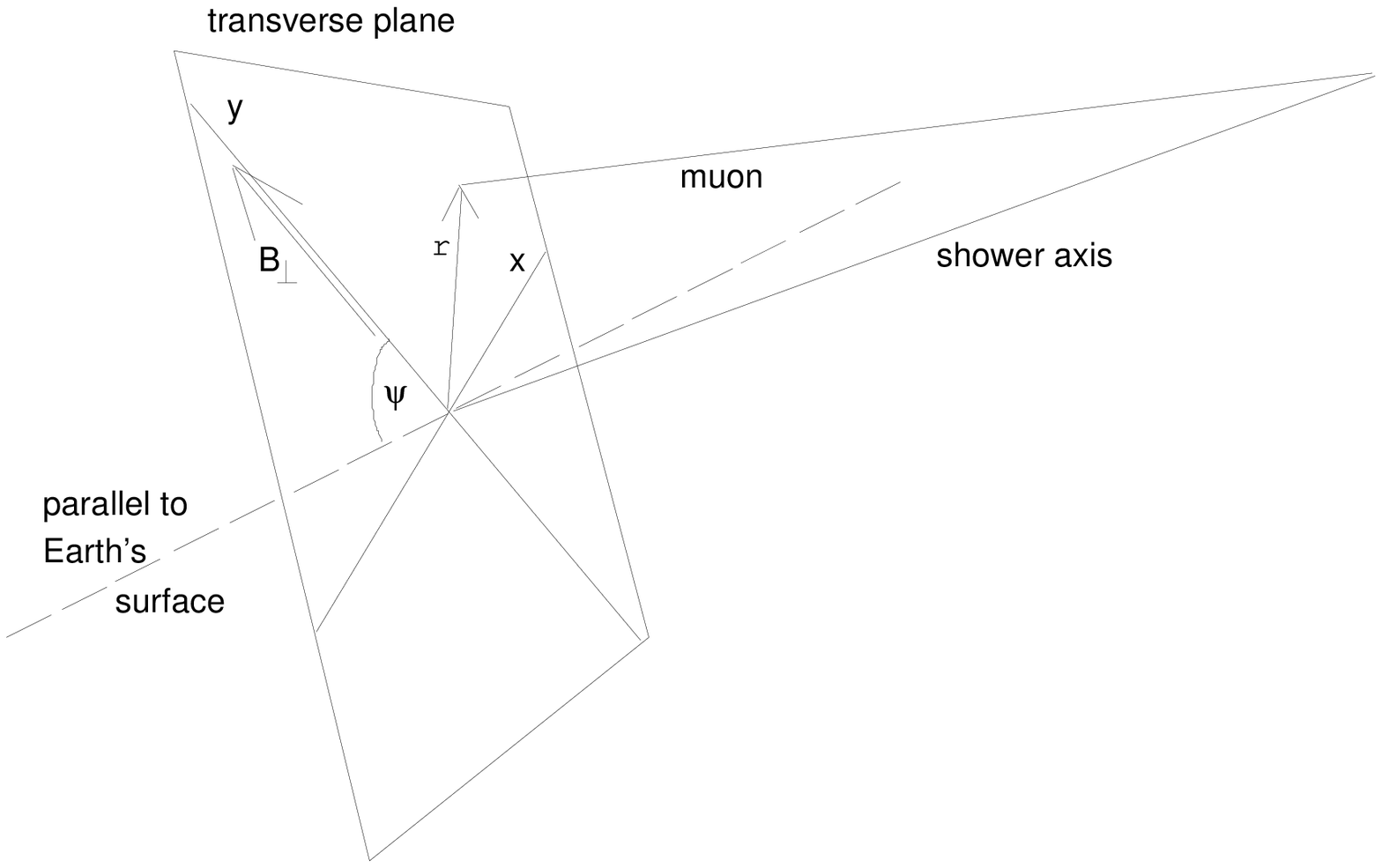,height=3.5in,width=3.5in}}
\vspace{10pt}
\caption{The plane perpendicular to the shower axis or 
{\sl transverse plane} and some useful geometrical definitions 
for Horizontal Air Showers. The $y$ axis is chosen along the direction of 
$B_{\perp}$ which is the projection of the magnetic field onto the 
transverse plane. $\psi$ is the angle subtended by the $y$ axis and 
the intersection of the transverse and the ground planes.}
\label{geometry}
\end{figure}

%
%
Consider a shower muon of an inclined shower produced high in the 
atmosphere at a distance $d$ to ground level and reaching ground level at 
a distance $r$ to shower axis in a plane transverse to the shower axis 
as schematically shown in Fig.~\ref{geometry}. 
This muon has an energy $E$ sufficiently high to reach ground level and 
let us assume that it has a transverse momentum $p_{\perp}$ that is 
small compared to $E$. In the absence of magnetic fields the muon travels 
to ground in a straight line. If we assume that the start of the muon track 
is right on the shower axis then clearly $r$ is related to $d$, 
$p_{\perp}$ and to $p$ (or $E$) through: 
\begin{equation}
r=  \frac{p_{\perp}}{p}~d \simeq \frac{c p_{\perp} }{E}~d.
\label{rE}
\end{equation}

For inclined showers both $p_{\perp}$ and $d$ 
in Eq.~\ref{rE} have distributions. It is not difficult to conclude that 
the $d$ distribution is narrow compared to the value of $d$. The 
majority of the muons are produced in a relatively narrow depth 
interval corresponding to the shower maximum for the parents (mostly 
pions that have average energy $25 \%$ larger than $E$). This range 
is of order a few radiation lengths, say 200~g~cm$^{-2}$ while the 
distance $d$ is controlled by the atmospheric slant depth from the 
production site to the ground which is over 3800 (10,000)~g~cm$^{-2}$ 
for showers with zenith angles above $75^{\circ}$ ($85^{\circ}$). 
The $p_{\perp}$ 
distributions on the other hand has an average of order 200~MeV/c. 
The value of $d$ sets the 
scale of the problem because it controls the matter depth from the 
production site to the ground and thus determines the minimum energy 
that a muon needs to reach the ground before losing its energy. 
Clearly as the zenith angle rises both $d$ and the average muon 
energy must also rise. The 
average values and the standard deviations of $d$ are shown in 
Table~\ref{alturas} together with the average muon energy at 
production and the average number of muons at ground level for 
different zeniths. 

%
%
\begin{table}
\begin{center}
\begin{tabular}{||c|c|c|c|c|c||} 
$\theta$ \mbox{(degrees)} & \mbox{d (km)} & \mbox{$\Delta$d (km)} & 
\mbox{ $<E>$ (GeV)} &  
$N_\mu \times 10^{-6}$& $\Delta N_\mu  \times 10^{-6}$ \\
\hline
$0^\circ$  & 3.9   & 2.8 & 8.1  &  29.  & 6.5  \\
$60^\circ$ & 16    & 6.5 & 18.9 &  13.3 & 2.4  \\
$70^\circ$ & 32    & 10  & 32.9 &  7.8  & 1.4  \\
$80^\circ$ & 88    & 17  & 77   &  3.3  & 0.7  \\
$87^\circ$ & 276   & 31  & 204  &  1.2  & 0.2  \\
\end{tabular}
\caption{Relevant parameters for muon production as obtained in 100 
proton showers of energy $10^{19}$ eV with a relative thinning of
$10^{-6}$ simulated with AIRES using SIBYLL 1.6 cross sections. 
Average values and RMS deviations for production altitude (d), 
muon energy at production ($<E>$), and total number of muons at 
ground level ($N_\mu$).}
\label{alturas}
\end{center}
\end{table}

If one takes $d$ and $p_{\perp}$ to be fixed Eq.~\ref{rE} implies 
that the energy 
spectrum and the lateral distribution of the muons are dependent. 
If either function is known the other can be deduced from it. 
The approximation of fixing $d$ and $p_{\perp}$, crude as it seems, 
can be used to reproduce the lateral distribution in a wide range 
of distances to the shower axis from the energy spectrum of the muons. 
One needs to fix the distance $d$ to the average values obtained in 
simulations and $p_{\perp}$ to a value $\simeq 200$~MeV/c. 

\begin{figure}[hbt]
\centering
\mbox{\epsfig{figure=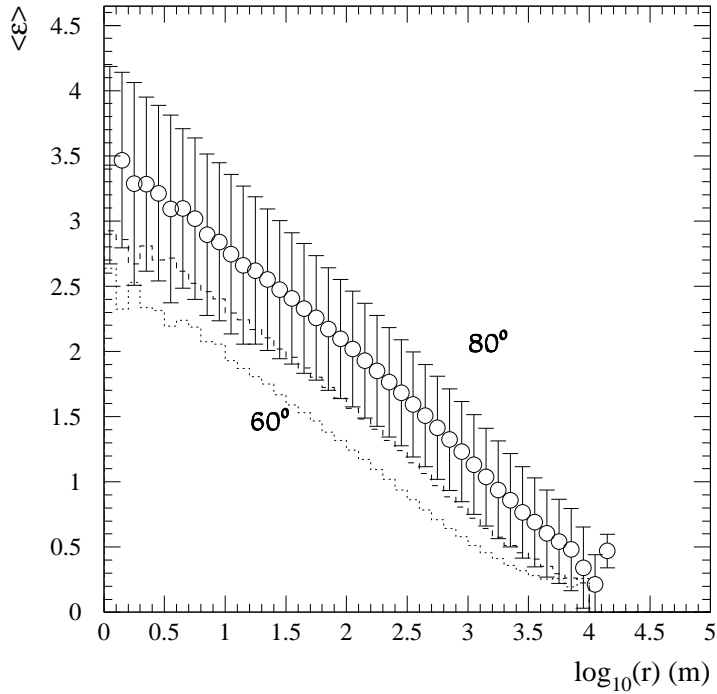,width=11.0cm}}
\caption{Correlation between $<\epsilon>=<\log_{10} E_\mu>$ and 
$\log_{10} r$ for zenith angles 60$^{\circ}$, 70$^{\circ}$, 
and 80$^{\circ}$ from bottom to top, see text. Error bars show the 
width of the $\epsilon$ distribution for 80$^\circ$.}
\label{epsilonfit}
\end{figure}

Eq.~\ref{rE} with fixed 
$p_{\perp}$ and $d$ also implies a particularly important 
anticorrelation between muon energy and distance to shower axis in 
the transverse plane. This anticorrelation shows up in 
the simulations of inclined showers very clearly as illustrated by the 
plot of the average muon energy as a function of $r$ shown in 
Fig.~\ref{epsilonfit}. The error bars in the graph indicate the standard 
deviations of the muon energy distributions. 

Deviations of the muon tracks by the Earth's magnetic field ($\vec B$) can be 
easily implemented by assuming that the muons are bent in a helix type 
trajectory which can be approximated by the arc of a circle. For small 
deviations only the projection of $\vec B$ onto the transverse plane, 
$\vec B_{\perp}$, matters. As a result it can be easily shown 
that the deviation of the muon in the transverse plane, $\delta x$, is at 
right angles to $\vec B_{\perp}$ and given by:
\begin{equation}
{\delta x}= R \left[1- \sqrt{1-\left(\frac{d}{R}\right)^2} \right] 
\simeq \frac {d^2} {2 R} = \frac{e B_{\perp} d}{2p},
\label{deltax}
\end{equation}
where $e$ is the electron charge and $p$ is the muon momentum 
and we have expanded brackets to first order. For $\delta x$ less than 
2~km we expect it to be valid at the $5 \%$ level at zenith $60^{\circ}$ 
because the typical production distance $< d >$ is 16~km. At higher 
zenith the approximation is excellent. 

When we combine the above relation with Eq.~\ref{rE} we obtain the 
following expression:
\begin{equation}
\delta x = \frac {e \vert B_\perp \vert d^2} {2p} = 
\frac{0.15 \vert B_\perp \vert d}{p_{\perp}} \; \bar r = \alpha \; \bar r,
\label{alpha}
\end{equation}
where in the last equation $B_{\perp}$ is to be expressed in Tesla, 
$d$ in m and $p_{\perp}$ in GeV. 
This expression is telling us that all positive 
(negative) muons that, in the absence of a magnetic field, would 
fall in a circle of radius $\bar r$ around shower axis, are translated a distance $\delta x$ to the right (left) of the $\vec B_{\perp}$ direction.   
The dimensionless parameter $\alpha$ measures the relative effect of the 
translation. When $\alpha << 1$ the magnetic effects are very small.  
However when $\alpha >1$ the magnetic translation exceeds the deviation 
the muons have due to their $p_{\perp}$. In this case {\sl shadow} 
regions with no muons 
are expected in the muon density profiles. For an approximate 
$p_{\perp} \sim 200$~MeV/c and $B_{\perp} = 40~\mu$T this happens when 
$d$ exceeds a distance of order 30~km, that is for zeniths above 
$\sim 70^{\circ}$.  
These shadow regions in the transverse plane are indeed an outstanding 
feature of the ground density profiles at high zeniths. 

It turns out that a very accurate description of the ground density profiles 
can be obtained using these simple ideas, provided one allows for an energy 
distribution of the muons at a fixed distance to shower axis, which has an 
average given by Eq.~\ref{rE}. Simulations have shown that a log normal 
distribution is adequate, or equivalently that $\epsilon=log_{10} E$ is 
assumed to be distributed with a gaussian of width $\sim 0.4$. 
The values of other inputs needed for calculating densities at each zenith 
angle such as the effective distance travelled by muons $d$, their lateral 
distribution function, and the average muon energy as a function of distance 
to shower axis are taken from simulations without magnetic field using 
AIRES \cite{AIRES}. 
It is a straightforward calculation to convert the radial distribution functions in the absence 
of a magnetic field to density patterns when the field is turned on, using 
the first equality in Eq.~\ref{alpha} and the corresponding coordinate 
transformation. Details of this method and extensive checks are given 
in Ref.\cite{HSmodel}. 
\begin{figure}[hbt]
\centering
\mbox{\epsfig{figure=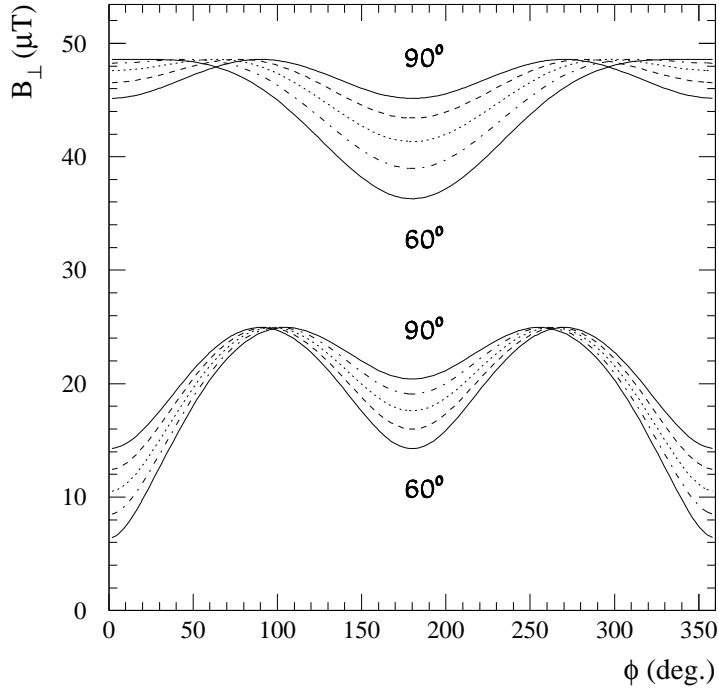,width=11.0cm}}
\caption{Magnetic field projection onto the transverse plane 
as a function of azimuthal angle for the Haverah Park 
location (top lines) and Pampa Amarilla (bottom lines). Lines are 
for zenith angles 60$^\circ$, 70$^\circ$, 80$^\circ$, 87$^\circ$, 
and 90$^\circ$.}
\label{Bmod}
\end{figure}
%

%
%

The particle densities in the transverse plane only need to be converted 
to the ground plane. It is remarkable that the relevant component of the 
magnetic field is its projection onto the transverse plane, $\vec B_{\perp}$. 
This projection in principle depends on the azimuthal direction of the 
shower and can be very different depending on the local orientation of 
the magnetic field vector, or equivalently on 
the array site location. Besides the magnitude of the magnetic field it is 
in fact the vertically upward component that is responsible for many of 
these quantitative differences between sites. The direction of 
$\vec B_{\perp}$ in the transverse plane changes as the azimuthal 
angle of the incident shower is changed. Depending on the site under 
consideration the 
changes in magnitude and direction of $\vec B_{\perp}$ as the 
incident azimuthal angle is varied can be very important. 
As an example this is illustrated in Fig.~\ref{Bmod} 
where the magnitude of $\vec B_{\perp}$ is plotted as a function of 
azimuth for two different site locations, that of Haverah Park and that 
of {\sl Pampa Amarilla} where the southern hemisphere Auger observatory 
is being constructed, more details can be found in \cite{HSmodel}. 

\section*{Characterization of inclined showers}

The approach described in the previous two sections has allowed a lot of  
progress in analysing existing data on HAS. Technically it allows the 
separation of azimuthal effects from other effects that are more interesting 
from the physics point of view. Systematic studies have been 
made of the changes in the transverse density patterns as the
primary energy, zenith angle and composition of the cosmic ray is changed 
and the results are also very enlightening. 

The effect of zenith angle has already been discussed. As the zenith angle 
rises the distance travelled by the muons $d$ and their energy $E$ rises. 
This makes inclined showers rather different from the vertical ones and, 
since $d$ changes by over a factor 10 between $60^{\circ}$ and $90^{\circ}$, 
it also implies that $85^{\circ}$ showers are very different from 
$70^{\circ}$ showers. As the zenith angle is 
increased the muons that are sampled at the ground have increasingly 
higher energies, and thus relate to earlier stages in shower development. 
Inclined showers are probing the shower core and the Earth's magnetic 
field is acting as an spectrometer for these high energy muons. 

\begin{figure}
\centerline{\epsfig{file=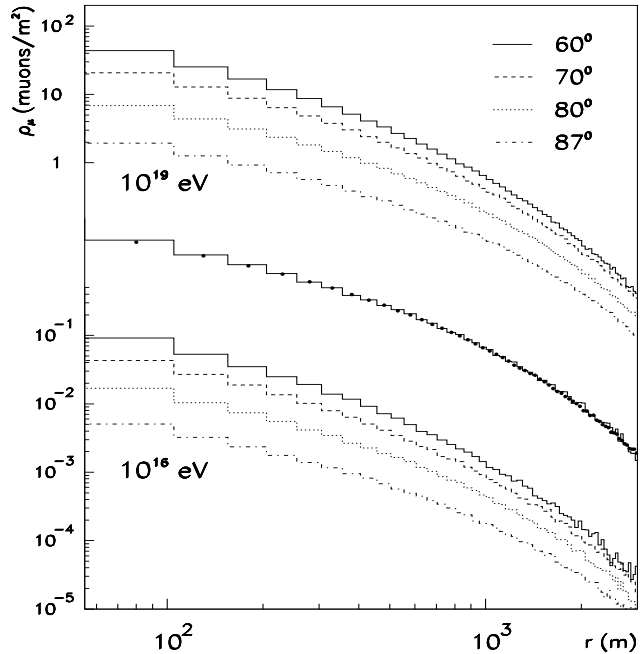,height=3.5in,width=3.5in}}
\caption{Lateral distribution functions for primary protons of 
energy 10$^{16}$, 10$^{17}$, 10$^{18}$ and 10$^{19}$ eV using 
SIBYLL.}
\label{LDF}
\end{figure}

Once the zenith angle is fixed there is little that changes from shower 
to shower and the effect is spectacular for showers of a given composition 
once the interaction model is fixed. 
As the energy of the primary proton changes, 
to an excellent approximation, the muon lateral distribution in the 
absence of a magnetic field has a universal 
shape. As a result it can be described by a single parameter, 
the normalization or equivalently the total number of muons in the shower. 
This effect is illustrated in Fig.~\ref{LDF} showing the lateral muon 
distribution for different zenith angles and a range of energies spanning 
the interval $10^{16}-10^{19}$~eV. The situation is analogous for the two 
hadronic models used, QGSJET \cite{qgsjet} and SIBYLL 
\cite{sibyll}. 

The result is not surprising after all. 
The scaling in showers has been known for a very long time and reflects 
the fact that the distributions are governed by the low energy physics that 
affects the majority of the shower particles. Besides the total number of 
particles, the main difference in showers of different energy is the 
position of shower maximum. It is well known that shower maximum only 
varies logarithmically with shower energy and this is a relatively small 
change when compared to the distance $d$ travelled by the muons in HAS. 
Once the muons are produced they are not very much affected until they 
reach ground level and as a result only the total number of muons produced 
matters to a very good approximation. This is in contrast to vertical 
showers where arrays are also measuring the electromagnetic contribution 
that, after reaching shower maximum becomes exponentially attenuated 
with depth and thus the densities observed at ground level are very 
sensitive to the position of shower maximum. 

For showers initiated by a heavy nucleus the situation is analogous and 
moreover the resulting lateral distribution is also extremely similar 
to that of protons. 
The explanation is the same since, besides the normalization, the main 
difference between the muons in proton and heavy ion showers is again 
the depth of maximum. 
\begin{figure}
\centerline{\epsfig{file=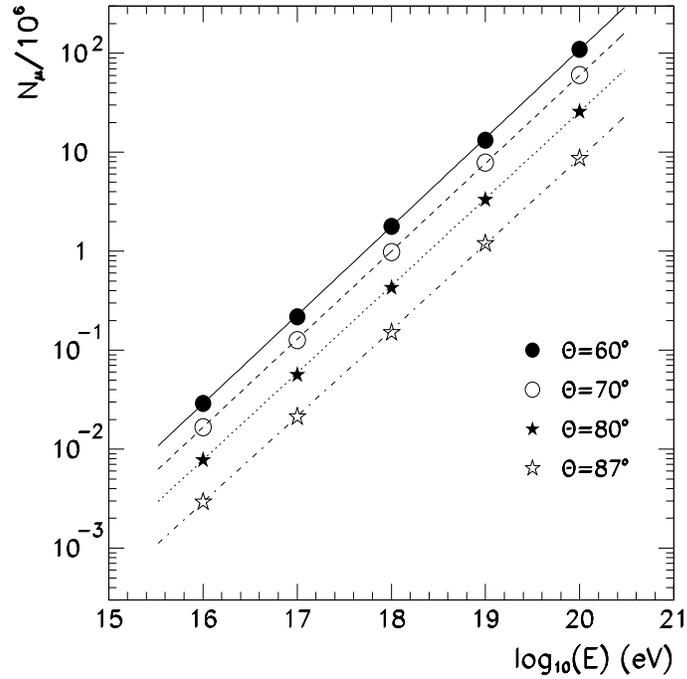,height=3.5in,width=3.5in}}
\caption{The relationship of total muon number to primary energy 
for protons of four zenith angles using the SIBYLL model.}
\label{scaling}
\end{figure}
For each zenith angle we thus characterize the shower by the total number 
of muons. The total number of muons in proton showers is shown in 
Fig.~\ref{scaling} as a function of primary energy for the SIBYLL 
hadronic model. The results can be easily parameterized by a function of the 
following type:
\begin{equation}
N=N_{\mu}~E^\beta
\label{Escaling}
\end{equation}
where $\beta$ is a constant for a given model and mass composition 
as shown in Table~\ref{nmu.tab}. 

In the case of photons the lateral distributions 
are also quite independent of primary energy with the exception of 
showers affected by the LPM effect. But the LPM effect is a density 
effect and is thus less important for horizontal than 
for vertical showers. Moreover the interactions of high energy photons 
with the Earth's magnetic field effectively imply that a high energy 
photon can be consider as a bunch of lower energy photons that are less 
likely to be affected by the LPM effect. The density profiles of the 
muons produced by photon primaries are similar to those 
produced by protons but the agreement is not as good as between protons 
and heavy nuclei because the dominant shower interactions are different. 
Photon showers 
are slightly narrower. To a relatively good approximation however the 
muon lateral distribution of photon showers can be assumed to be like 
that of protons with $\beta=1.2$ and with 
$N_{\mu}(10^{19}~{\rm eV})=5.7~10^5$. Although the rise of the number of 
muons with energy is more rapid for photon primaries than for proton, 
at $10^{19}$~eV photons are still 9 times poorer in muons than protons. 

\begin{table}
\begin{center}
\begin{tabular}{|lrcc|} \hline
Model & A & $\beta$ & $N_{\mu}$ ($10^{19}$ eV) \\\hline\hline
SIBYLL & 1 & 0.880 & 3.3 10$^{6}$ \\
       & 56 & 0.873 & 5.3 10$^{6}$ \\ \hline
QGSJET & 1 & 0.924 & 5.2 10$^{6}$ \\
       & 56 & 0.906 & 7.1 10$^{6}$ \\ \hline
\hline
\end{tabular}
\end{center}
\caption{Relationship between muon number and primary energy for 
two hadronic models and two primary masses (see equation~\ref{Escaling}).}
\label{nmu.tab}
\end{table}

\section*{Interpretation of HAS data}

The ability to generate muon density profiles makes it possible to 
make rate simulations for a given array geometry and in particular 
that of Haverah Park with which nearly 10,000 events were recorded 
with zenith angles above $60^{\circ}$ between 1974 and 1987. 
The Haverah Park detector was 
a 12~km$^2$ air shower array using 1.2~m deep water \v Cerenkov 
tanks in Northern England and 
which has been described elsewhere \cite{haverah}. 
For the study to be described the recorded data for zenith angle 
above $60^{\circ}$ were 
reanalyzed for arrival direction for the described analysis using 
all the time information available for the tanks. This gave an 
improvement in angular accuracy. 

The process of 
making this simulation is however rather elaborate because the signals 
in the detectors have to be carefully calculated. It is clear that 
signals in water \v Cerenkov tanks from muons close to the horizontal 
directions are rather different from vertical incidence. The shower 
muons are more energetic than in vertical showers and thus more likely 
to produce interactions 
in the tank which complicate the signal distributions. These corrections 
to the signal are discussed in detail in Ref.~\cite{rate} where it is 
shown that 
the corrections due to delta rays, direct light going into the 
phototubes, the electromagnetic component due to muon decay and muon 
interaction processes in the tank (mostly pair production and 
bremsstrahlung) are of great importance and increase the rate by 
a factor between 3 and 4 with respect to a calculation using only the 
signal due to the muon tracks. The use of signal distributions becomes 
essential and these were calculated with WTANK \cite{Wtank}, a program 
based on GEANT.  

The measured trigger rate as a function of zenith angle is well 
reproduced in magnitude and azimuth when these corrections 
are carefully taken into account. The rate simulation can be 
made assuming different composition and/or different hadronic 
models and the results compared to the data. While protons 
slightly underestimate the rate, heavy nuclei overestimate it, so 
clearly a mixture of the two can reproduce the data for any of the 
two hadronic interaction models considered \cite{rate}. 

The fact that smooth average muon density profiles 
can be obtained with the described prescription allows the possibility 
of making fits to the data in an attempt to evaluate the energy and 
impact parameter of individual showers once a given composition is 
assumed. For zeniths below $70^{\circ}$ a remnant electromagnetic 
contribution from the showering process is considered, 
which is concentrated near the shower core. 
The fitting process is complex because the Haverah Park array at 
large zenith angles only samples a small band in the transverse 
plane of the shower and uncertainties in arrival directions as well 
as correlations between fitted parameters have to be taken into 
account. To improve the fits additional information from 
an infilled portion of the array \cite{infill} that was running 
simultaneously part of the time was included when available. 

\begin{figure}
\centerline{\epsfig{file=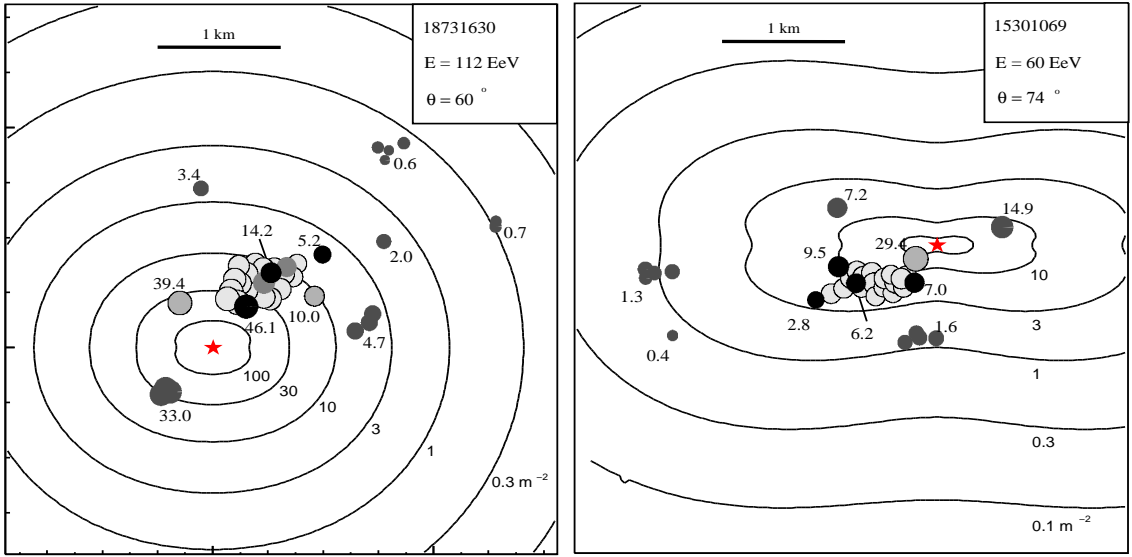,height=3.in,width=6.in}}
\caption{Density maps of two events in the plane perpendicular 
to the shower axis. Recorded muon densities are shown as circles 
with radius proportional to the logarithm of the density. The detector
areas are indicated by shading; the area increases from white to black as
1, 2.3, 9, 13, 34 m$^{2}$. The position of the best-fit core is
indicated by a star. Selected densities are also marked. The y-axis 
is aligned with the component of the magnetic field perpendicular to the
shower axis.}
\label{events.fig}
\end{figure}

The correlation between arrival direction and energy demands an 
iterative process in which the arrival directions are corrected once 
the core position is determined. This refitting procedure for the 
arrival directions takes into consideration both the curvature 
corrections and the distribution of the arrival time of the first 
muon in terms of the signal received in the tanks. Once the first 
density fit is performed with the arrival directions obtained with 
a time fit to a plane front, the sequence of a time fit with 
curvature corrections and a density fit with the new arrival 
directions is repeated three times. Two examples of the results 
of such fits compared to the average density contours are shown 
in Fig.~\ref{events.fig}.

Once the data have been analysed a number of quality cuts are performed 
to eliminate badly reconstructed events. A cut has been taken 
selecting showers with shower core at a distance less than 2~km from 
the array center to make sure that the event is well contained within the 
array. A second cut has been made on the density fit rejecting those 
events that have a $\chi^2$ probability below $1\%$. Lastly a cut has 
been made to reject events that have a large error in the energy 
determination which is defined to be the sum in quadrature of the 
error from the fit and the error due to the uncertainty in the 
arrival direction. This error is required 
to be below $50 \%$: this automatically eliminates all showers with zenith 
angles above $80^{\circ}$. After these cuts are performed 2, 7 and 46 
events are reconstructed with proton equivalent energies above 
$10^{20}~$eV, $4~10^{19}~$eV and $10^{19}~$eV, respectively. 

The procedure is applied to the data and to simulation using the cosmic ray spectrum of reference \cite{WatsonNagano}. The simulated data goes through an 
identical fitting procedure and the resulting rate is compared to 
measurement. The agreement between the integral rate above $10^{19}$~eV 
measured and that obtained with simulation is striking when 
the QGSJET model is used. Sibyll leads to a slight underestimate of the 
observed rate. 

In the context of this conference what is of more importance is the new 
possibility that HAS open for studying composition at the highest 
energies. The universality of the muon lateral distribution function 
is very powerful and once the equivalent proton energy is determined 
for all events, the corresponding energies under the assumption 
that the primaries are iron nuclei (photons) can be obtained by 
multiplying the proton energy 
by a factor which is $\sim 0.7$ (6) for $10^{19}$~eV and which 
varies slowly with equivalent proton energy.  
As a result when a photon primary spectrum is 
assumed the simulated rate seriously underestimates the observed data 
by a factor between 10 and 20. A fairly robust bound on the 
photon composition at ultra high energies can be established assuming a
two component proton-photon scenario. The photon 
component of the integral spectrum above $10^{19}~$eV (4~$10^{19}$~eV) 
must be less than $41\%$ ($65\%$) at the $95\%$ confidence level. 
Details of the analysis are presented in \cite{avePRL} and will be 
expanded elsewhere. A similar analysis can be made for a two component 
scenario with protons and iron nuclei. When the QGSJET model is used a 
similar limit restricting the iron component to less than $54\%$ 
above $10^{19}$~eV can be similarly established. In this 
case the iron only assumption leads to an overestimate of the 
rate. The limit obtained is however less robust in the sense that it is 
fairly sensitive changes in the interaction model and/or assumptions 
about the incident cosmic ray flux \cite{avePRL}. 

\section*{Conclusions}

I have summarized recent progress in understanding density 
patterns of Extensive Air Showers at high zenith angles 
performed by the Leeds-Santiago collaboration and which 
is described in detail in references \cite{avePRL,HSmodel,rate}. 

I have reviewed a scheme for understanding the complex muon density 
patterns that develop at ground level because of the geomagnetic 
effects on the muon component of these showers. 
The scheme can be used as an effective parameterization of the 
average density profiles at ground level and simplifies the 
task of performing simulations to be compared with data. 

The ideas have been successfully tested with the Haverah Park 
data and the measured rate is consistent with simulations. 
The recorded individual density patterns 
can be fitted to average values obtained in this scheme and 
the energy of the cosmic rays extracted for a given composition. 
The analysis of inclined air shower data from Haverah Park Array 
for energies above $10^{19}$~eV has been shown to be consistent 
with a proton composition using the QGSJET model for hadronic 
interactions. As a result of this study the photon component of 
the UHECR spectrum at energies above $10^{19}~$eV must be less 
than $41\%$ with a $95\%$ confidence level. 

The detection of large HAS provides a new tool for the study of 
high energy cosmic ray which has been shown to be very sensitive 
to primary composition in the EeV range and above. The 
cosmic ray spectrum is measured by the detection of vertical showers 
in a composition independent fashion. For high zeniths 
particle arrays detect mostly intermediate energy muons at ground 
level and the measured rate becomes particularly sensitive to 
photon composition on the basis of the reduced muon content of 
photon induced showers. The analysis of such inclined showers can 
effectively double the aperture of any given extensive air shower 
detector array at these energies and particularly for the  
Auger Observatories now in costruction. The prospects for 
establishing photon composition at EeV energies and above by  
such arrays are significantly enhanced using this method.  

\vskip 0.5 cm
{\bf Acknowledgements:} I thank A.A.~Watson and R.A.~V\'azquez for 
suggestions after reading this manuscript. 
This work was supported in part by CICYT (AEN99-0589-C02-02) and by 
Xunta de Galicia (PGIDT00PXI20615PR). 
 

\begin{references}
%
\bibitem{Tokyo} T.~Hara {\sl et al.} 
{\sl Proc. of the XI Int. Cosmic Ray Conf.}, Budapest (1969), 
Acta Physica Academiae Scietiarum Hungaricae 29, Suppl. 3, p. 125-131, (1970); Nagano,~M. {\sl et al.}, {\sl J.\ Phys. Soc. Japan}
{\bf 30} 33 (1971).
%
\bibitem{Kiel} E.~B\"ohm {\sl et al.}, 
{\sl Proc. of the XI Int. Cosmic Ray Conf.}, Budapest (1969), 
Acta Physica Academiae Scietiarum 
Hungaricae 29, Suppl. 3, p. 121-124, (1970).
%
\bibitem{Durham} Alexander,~D. {\sl et al.}, 
{\sl Proc. of the XI Int. Cosmic Ray Conf.}, Budapest (1969), 
Acta Physica Academiae Scietiarum 
Hungaricae 29, Suppl. 3, p. 215, (1970).
%
\bibitem{Andrews} D. Andrews {\sl et al.}
{\sl Proc. of the XI Int. Cosmic Ray Conf.}, Budapest (1969), 
Acta Physica Academiae Scietiarum 
Hungaricae 29, Suppl. 3, p. 337-342, (1970).
%
\bibitem{markov} M.A.~Markov, in {\sl Proc. of the Annual Conf. on 
High Energy Physics}, Rochester (1960), p. 578.
%
\bibitem{berezinskii} V.S.~Berezinsky and G.T.~Zatsepin, 
{\sl Yad.\ Fiz.\ } {\bf 10} (1969) 1228. [{\sl Sov.\ J. Nucl.\ Phys} 
{\bf 10} (1969) 696]. 
%
\bibitem{Kiraly} P. Kiraly {\it et al.}, {\sl J. Phys. A: Gen. Phys.} 
{\bf 4} (1971) 367.
%
\bibitem{tokiobremss} S.~Mikamo {\em et al.}, {\sl Lett.\ al Nuovo Cimento\ 
{\bf 34} N 10, (1982) 273}. 
%
\bibitem{hsvum} E.~Zas, F.~Halzen and R.A.~V\'azquez, Astropart.\ Phys.\  
{\bf 1}, 297 (1993). 
%
\bibitem{Antonov} Yu.N.~Antonov, Yu.N.~Vavilov, G.T.~Zatspin {\sl et al.},
{\sl Zh. \'Eksp. Teor. Fiz.} {\bf 32} (1957) 227 ({\sl Sov. Phys. JETP} 
{\bf 5} (1957) 172).  
%
\bibitem{Hillas} A.M. Hillas {\sl et al.}, 
{\sl Proc. of the XI Int. Cosmic Ray Conf.}, Budapest (1969), 
Acta Physica Academiae Scietiarum 
Hungaricae 29, Suppl. 3, p. 533-538, (1970).
%
\bibitem{nagano86} Nagano,~M. {\sl et al.}, 
{\sl J.\ Phys. G: Nucl. Phys.} {\bf 12}  69 (1986); 
%
\bibitem{baltrusaitis} R.M. Baltrusaitis {\it et al.}, Phys. Rev. 
{\bf D 31}, 2192 (1985).
%
\bibitem{nureview} For a review see for instance, P. Bhattacharjee and 
G. Sigl, {\sl Phys.\ Rept.\ } {\bf 327} (2000) 109-247, 
T.K. Gaisser, F. Halzen and T. Stanev, {\sl Phys.\ Rept.\ } {\bf 238} 
(1995) 173, and references therein.
%
%
\bibitem{hznubound} F.~Halzen and E.~Zas, Phys. Lett. {\bf B289} (1992) 
184-188; F.~Halzen and E.~Zas, In {\sl Proc. of the High energy neutrino 
astrophysics}, Honolulu, World Sci., Singapore (1992), p 186-195.   
%
\bibitem{stecker91}  F.W.~Stecker, C.~Done, M.H.~Salamon and P.~Sommers,
{\sl Phys.\ Rev.\ Lett.\ } {\bf 66} (1991) 2697; F.W.Stecker,  M.H.Salamon 
{\sl Space\ Sci.\ Rev.} {\bf 75} (1996) 341-355
%
\bibitem{Bhattacharjee} P. Bhattacharjee {\it et al.}, {\sl Phys. Rev. Lett.}
{\bf 69} (1992) 567.
%
\bibitem{Weiler}  T.J. Weiler, Astropart. Phys. {\bf 11}, 303 (1999). 
%
\bibitem{Fargion} D. Fargion, B. Mele, and A. Salis, Astrophys. J. 
{\bf 517}, 725 (1999). 
%
\bibitem{BlancoPRL} J.J. Blanco-Pillado, R.A. V\'azquez, and E. Zas, 
Phys. Rev. Lett. {\bf 78}, 3614 (1997).
%
\bibitem{BlancoPRD} J.J. Blanco-Pillado , R.A. Vazquez, E. Zas,  
Phys. Rev. {\bf D61} (2000) 123003. 
%
\bibitem{klypve} B. Khrenov in these proceedings. 
%
\bibitem{euso} O. Catalano and L. Scarsi in these proceedings. 
%
\bibitem{Auger} {\sl The Pierre Auger Project Design Report}.
By Auger Collaboration. FERMILAB-PUB-96-024, Jan 1996. 252pp.
%
\bibitem{venice}
G.~Parente and E.~Zas, in {\sl Proceedings of the 7th Int.Symposium on 
Neutrino Telescopes.} p.~345, ed.\ by M.~Baldo Ceolin, Venice
(1996).
%
\bibitem{Capelle} J. Capelle, J.W. Cronin, G. Parente, and 
E. Zas, {\sl Astropart. Phys.} {\bf 8} (1998) 321.
%
\bibitem{avePRL} M. Ave, J.A. Hinton, R.A. Vazquez, A.A. Watson, 
and E. Zas, Phys. Rev. Lett. {\bf 85}, (2000) 2244.
%
\bibitem{hillasrho} A.M. Hillas, D.J.~Marsden, J.D.~Hollows, and 
H.W.~Hunter, in {\sl Proc. of the XII Int. Cosmic Ray Conf.}, 
Hobart {\bf 3} (1971) 1001. 
%
\bibitem{Antonov98} E.E.~Antonov, L.G.~Dedenko, Yu.P.~Pyt'ev {\sl et al.},
{\sl Pis'ma Zh. \'Eksp. Teor. Fiz.} {\bf 68} (1998) 177 ({\sl JETP Letts.} 
{\bf 68} (1998) 185).   
%
\bibitem{Ivanov} A.A.~Ivanov {\sl et al.}, {\sl Pis'ma Zh. \'Eksp. Teor. Fiz.} {\bf 69} (1998) 263 ({\sl JETP Letts.} {\bf 69} (1999) 288). 
%
\bibitem{AIRES} AIRES: A System for Air Shower Simulation, S.J.~Sciutto, 
{\sl Proc. of the XXVI Int. Cosmic Ray Conf.} Salt Lake City (1999), 
vol. 1, p.~411-414; S.J.~Sciutto, {\sl preprint archive} astro-ph/9911331.   
%
\bibitem{HSmodel} M.~Ave, R.A.~V\'azquez, and E.~Zas, 
{\sl Astropart.\ Phys.\ } 14 (2000) 91.
%
\bibitem{qgsjet} N.N. Kalmykov and S.S. Ostapchenko, {\sl Yad. 
Fiz.} {\bf 56} (1993) 105; {\sl Phys. At. Nucl.} {\bf 56}(3) 
(1993) 346; N.N. Kalmykov, S.S. Ostapchenko, and A.I. Pavlov, 
{\sl Bull. Russ. Acad. Sci. (Physics)} {\bf 58} (1994) 1966.
%
\bibitem{sibyll} R.T. Fletcher, T.K. Gaisser, P. Lipari, and 
T. Stanev, {\sl Phys. Rev.} {\bf D50} (1994) 5710; J. Engel,
T.K. Gaisser, P. Lipari, and T. Stanev, {\sl Phys. Rev.} {\bf D46}
(1992) 5013.
%
\bibitem{haverah} R.M. Tennent, {\sl Proc Phys Soc} {\bf 92} 
(1967) 622. M.A. Lawrence, R.J.O. Reid, and A.A. 
Watson, {\sl J Phys G} {\bf 17} (1991) 733. 
\bibitem{rate} M. Ave, J.A. Hinton, R.A. Vazquez, A.A. Watson, 
and E. Zas, Astropart. Phys. 14 (2000) 109.
%
\bibitem{Wtank} J.R.T. de Mello Neto, WTANK: A GEANT Surface Array 
Simulation Program GAP note 1998-020. 
%
\bibitem{infill} D M Edge  {\sl et al.} 
{\sl Proc. of the XV Int. Cosmic Ray Conf.} (Plovdiv) vol. 9 (1977) p. 137.
%
\bibitem{WatsonNagano} M.~Nagano and A.A.~Watson, {\sl Rev. Mod. Phys.} 
72 (2000) 689. 
%
\end{references}
\end{document}